\documentclass[a4paper,11pt]{article}
\usepackage{epsfig,cite,mathbbol,graphicx}

\usepackage{amsmath}
\usepackage{amssymb}
\usepackage{a4wide}
\usepackage{fancyhdr}
\usepackage{macros}

\newcommand{\Nr}{N_{\rm{r}}}

\begin{document}

\begin{titlepage}
\begin{flushright}
{\small\tt
HIM-2011-10 \\
IFT-UAM/CSIC-11-90\\
CERN-PH-TH-2011-296\\
MKPH-T-11-19}
\end{flushright}

\vskip 0.8 cm
\begin{center}
  {\Large\bf On the efficiency of stochastic volume sources for the
             determination of light meson masses \\[0.5ex]}
\end{center}
\vskip 0.5 cm
\begin{center}
{\large E. Endress,$^{1}$ A. J\"{u}ttner,$^{2}$ and H. Wittig$^{3}$
}
\vskip 0.5cm
$^{1}$Instituto de
  F\'{i}sica The\'{o}rica UAM/CSIC, Facultad de Ciencas, Universidad
  Aut\'{o}noma de Madrid, Cantoblanco, E-28049 Madrid, Spain
\vskip 1.5ex
$^{2}$CERN, Physics Department, TH Division, CH-1211 Geneva 23,
  Switzerland
\vskip 1.5ex
$^{3}$Institut f\"{u}r Kernphysik and Helmholtz Institute Mainz,
  Johannes Gutenberg-Universit\"{a}t Mainz, D-55009 Mainz, Germany
\vskip 1.0cm
{\bf Abstract}
\vskip 0.35ex
\end{center}

\noindent

We investigate the efficiency of single timeslice stochastic sources
for the calculation of light meson masses on the lattice as one varies
the quark mass. Simulations are carried out with $N_{f}=2$ flavours of
non-perturbatively $\mathcal{O}(a)$ improved Wilson fermions
for pion masses in the range of $450-760\,\MeV$. Results for
pseudoscalar and vector meson two-point correlation functions computed
using stochastic as well as point sources are presented and compared.
At fixed computational cost the stochastic approach reduces the
variance considerably in the pseudoscalar channel for all simulated
quark masses.
The vector channel is more affected by the intrinsic stochastic noise. In
order
to obtain stable estimates of the statistical errors and a more pronounced
plateau for the effective vector meson mass, a relatively large number of
stochastic sources must be used.

\vfill

\begin{center}
November 2011
\end{center}

\eject

\vfill
\eject

\end{titlepage}

\section{Introduction}

In lattice QCD, hadronic properties such as masses and matrix elements
can be computed in terms of Euclidean correlation
functions. Typically, these are expectation values of properly
selected polynomials in the quark and gluon fields that project on
states with the desired quantum numbers. After performing the Wick
contractions, correlation functions are expressed as traces over
products of quark propagators, Dirac matrices and color-structures.
In order to obtain precise estimates of hadron masses and transition
amplitudes, variance reduction methods may be applied such that
correlation functions with good statistical accuracy can be computed.

The quark propagator in coordinate space can be computed as the
solution of the linear system
\begin{equation}
   D\Phi = \eta,
\label{eq:linsyst}
\end{equation}
where $D$ is the lattice Dirac operator and $\eta$ a source
vector. In its simplest form, $\eta$ is taken to be a point source,
i.e.\footnote{For simplicity we suppress colour and spinor indices.}
\begin{equation}
  \eta(x^\prime) = \delta_{x^{\prime}y}\,.
\end{equation}
This implies that the solution of eq.\,(\ref{eq:linsyst}) yields the
quark propagator from a single point~$y$ to any other point~$x$, which
corresponds to just one column of the propagator matrix. Thanks to
translational invariance a large class of correlation functions can be
defined in terms of these ``one-to-all'' propagators. However, in
this way only a small fraction of the information contained in $D$ is
processed. In typical simulations of lattice QCD the sparse matrix $D$
has $O(10^9\times 10^9)$ entries, and solving eq.\,(\ref{eq:linsyst})
to machine precision for all source positions is therefore beyond the
capabilities of even the most powerful supercomputers. Volume-filling
random-noise sources have been proposed as a means to access the full
propagator matrix~\cite{Bernardson,Dong_Liu,RomeII,Peisa} by replacing
it with a stochastic estimate.
%
%
%
These stochastic ``all-to-all'' propagators have been successfully applied in
a number of different
contexts\,\cite{Dong_Liu,RomeII,Peisa,Foster,SESAM,Cais,Foley,McNeile,
  Simula,andreas_article,ETM,ETMeta}. However, care must be exercised in their
actual construction, since an arbitrarily chosen stochastic source vector can
induce a large variance into hadronic correlation functions. In other words,
for the method to be efficient, one must ensure that the intrinsic stochastic
noise does not overwhelm the gain in information provided by having access to the entire
propagator matrix. A particular technique, which proved to be quite efficient,
is the so-called ``one-end trick'', pioneered in \cite{Foster,McNeile} and
successfully applied in several studies of light meson
physics\,\cite{Simula,andreas_article,ETM,ETMeta}.

In this article we present a systematic study of the effectiveness of
single-timeslice stochastic sources. In particular, we monitor the
variance and the computational cost as a function of the quark mass
and compare it to the variance achieved with point sources. Our
simulations are performed for QCD with $N_{\rm{f}}=2$ flavours of
$O(a)$ improved Wilson quarks. We concentrate on two-point correlation
functions of the pion and the rho meson.

The outline of this paper is as follows: in section\,\ref{theory} we
revisit the concepts of stochastic volume sources and the one-end
trick. The set-up of our simulations is discussed in
section\,\ref{Simulation_Setup}, and our main results are presented in
section\,\ref{results}. Finally, section\,\ref{summary} contains a
summary of our findings as well as some concluding remarks.

\section{Stochastic propagator estimation}\label{theory}

In this paper we restrict ourselves to two-point correlation functions
of a flavour off-diagonal quark bilinear, $O_{ud}(x)=\bar u(x)\Gamma
d(x)$, where $u$ and $d$ denote the fields of the up- and the
down-quarks, respectively, and $\Gamma$ is one of the 16 Dirac
matrices specifying the desired quantum number. The two-point
correlation function is given by
\begin{equation}\label{C2}
\left\langle
\mathcal{O}_{ud}(x)\mathcal{O}_{ud}^{\dag}(y)
        \right\rangle=\left\langle \text{Tr}\left\lbrace
	S(x,y)\Gamma S(y,x)
        \tilde{\Gamma}\right\rbrace \right\rangle,
\end{equation}
where $\tilde{\Gamma}=\gamma_{0}\Gamma^\dagger\gamma_{0}$, and
$\left\langle \dots \right\rangle$ denotes the average over the gauge
configurations. Since we assume exact isospin symmetry, the same
symbol, $S$, is used to denote the propagator for both quark flavours.
The spectrum of the particles with the prescribed quantum numbers
can then be determined from the exponential decay of
the zero-momentum projection, i.e.
\begin{equation}
C_{2}(t;y)=\sum_{\vec{x}}
        \left\langle \text{Tr}\left\lbrace S(x,y)\Gamma
        S(y,x)\tilde{\Gamma}\right\rbrace \right\rangle, \quad
        t=x_{0}-y_{0}. 
\end{equation}
For a given gauge field the propagator $S(x,y)$ from lattice sites $y$
to $x$, is obtained as the solution of the linear system\footnote{
Latin indices are used to label color, while Greek letters denote
spinor components.}
\begin{equation}
\label{eqn:source_method}
 \sum_{z}\sum_{c,\gamma}D_{\alpha\gamma}^{ac}(x,z)
        S_{\gamma\beta}^{cb}(z,y)= 
        \delta_{\alpha\beta}\delta^{ab}\delta_{xy}\,.
\end{equation}
Using point sources amounts to solving the linear system for one
particular choice of~$y$. This implies that altogether 12 inversions
must be performed, since four spinor and three colour components must
be considered independently. While this is sufficient to compute the
correlation function $C_2(t;y)$, the statistical signal could be
further improved by averaging over many different source points. An
improved estimator for the correlation function is obtained by
\begin{equation}\label{final_2pt}
\widetilde{C}_2(t) \equiv \frac{1}{V_3}\sum_{\vec{y}}C_{2}(t;y)=
        \frac{1}{V_3}\sum_{\vec{x},\vec{y}} 
        \left\langle \text{Tr}\left\lbrace S(x,y)\Gamma
        S(y,x)\tilde{\Gamma}\right\rbrace \right\rangle, \quad
        t=x_{0}-y_{0},
\end{equation}
where $V_3$ denotes the number of lattice sites within one timeslice.
In order to see how such an average can be effected, we revisit the
method of stochastic noise sources in the following.

\subsection{General stochastic noise sources}

In the stochastic approach, an ensemble of $\Nr$ random vectors,
$\left\lbrace \eta^{(r)}(x_0,\vec{x})|r=1,\dots,N_{r}\right\rbrace$,
is generated for each gauge configuration. The number $\Nr$ is
sometimes referred to as the number of ``hits''. These source vectors
are created by assigning independent random numbers to all components,
i.e. to all lattice sites, colour and Dirac indices. Each random
number is drawn from a distribution $\mathcal{D}$ which is symmetric
about zero in the hit limit ${N_{r}\rightarrow \infty}$, i.e.
\begin{equation}\label{symmetric_about_zero}
  \left\langle \eta^{a}_{\alpha}(x_0,\vec{x}) \right\rangle_{\rm{src}}
  \equiv \lim_{\Nr\to\infty} \frac{1}{\Nr}\sum_{r=1}^{\Nr}
  \big(\eta^{(r)}\big)_{\alpha}^{a}(x_{0},\vec{x})=0. 
\end{equation}
In addition the sources satisfy
\begin{equation}\label{orthonormality_condition}
  \left\langle \eta_{\alpha}^{a}(\vec{x},x_{0})
  (\eta^{\dag})_{\beta}^{b}(\vec{y},y_{0})
  \right\rangle_{\rm{src}} = 
  \delta_{x_{0}y_{0}}\delta_{\vec{x}\vec{y}}
  \delta_{\alpha\beta}\delta^{ab}.
\end{equation}
Solving the linear system of eq.~(\ref{eq:linsyst}) for each of the
$\Nr$ source vectors yields a set of solution vectors
\begin{equation}
  \big(\Phi^{(r)}\big) _{\alpha}^{a}(x)= \sum_{z}\sum_{c,\gamma}
  S_{\alpha\gamma}^{ac}(x,z) \big(\eta^{(r)}\big)_{\gamma}^{c}(z).
\end{equation}
The estimate of the entire propagator is defined as the stochastic
average (``hit average'') over the product between solution and noise
vectors
\begin{equation}\label{estimate}
  \left\langle \Phi_{\alpha}^{a}(x)(\eta^{\dag})_{\beta}^{b}(y)
  \right\rangle_{\rm{src}}
 = \sum_{z}\sum_{c,\gamma}
    S_{\alpha\gamma}^{ac}(x,z)\,\delta_{zy} \delta_{\gamma\beta}\delta^{cb}
 = S_{\alpha\beta}^{ab}(x,y).
\end{equation}
It remains to specify the distribution from which the random vectors
$\eta^{(r)}$ are drawn. In refs.\,\cite{Bernardson,Dong_Liu} it was
noted that a flat distribution of $\mathbb{Z}(2)$-elements,
$\left(\eta^{(r)}\right)_{\alpha}^{a}(x)\in
\mathcal{D}=\mathbb{Z}(2)=\left\lbrace +1,-1\right\rbrace $, or, more
generally, elements of $\mathcal{D}=\mathbb{Z}(N)$ is very effective
in realising the condition of
eq.\,(\ref{orthonormality_condition}). In this work we follow Foster
and Michael\,\cite{Foster} and draw separate elements of
$\mathbb{Z}(2)$ for the real and imaginary parts of the source vector,
i.e. 
\begin{equation}
   \mathcal{D}=\mathbb{Z}(2)\otimes \mathbb{Z}(2)=\left\lbrace
   \frac{1}{\sqrt{2}}\left(\pm1 \pm  i \right)  \right\rbrace.
\end{equation}
Experience shows that a random source vector, which is distributed
over the entire space-time lattice, leads mostly to a very noisy
signal for hadronic correlation functions. An essential step towards a
significant variance reduction is taken by restricting the support of
the source vector to individual timeslices, Dirac or colour components
\cite{Foley}. Such ``dilution schemes'', and, in particular, the
so-called time-dilution are widely used in the computation of hadronic
properties\,\cite{Foster,Cais,Foley}. Here, the non-zero components of
the random source vector are restricted to a single timeslice,
e.g. $y_0=0$
\begin{equation}
  \eta(y) = \left\{ \begin{array}{ll}
  \widetilde\eta(\vec{y}),\quad & \hbox{if}\; y_0=0, \\
  0,\quad & \hbox{otherwise} \end{array} \right.\,.
\end{equation}
We end this discussion with a few general remarks. In practice the
limit $\Nr\to\infty$ cannot be taken, and thus the details of
constructing the random source should be optimized for the correlation
function at hand. In ref.\,\cite{andreas_article} it was noted that
for many correlators the stochastic and gauge averages commute. It
then suffices to generate a reasonably small number of random source
vectors per gauge configuration, since the ensemble average
automatically implies the stochastic average.

\subsection{The one-end trick revisited}\label{sec:one-end-trick}

The naive implementation of stochastic sources consists in replacing each
propagator in the correlation function of eq.\,(\ref{final_2pt}) by the
stochastic estimate of eq.\,(\ref{estimate}). For a general choice of $\Gamma$
this implies that an independent source vector must be used for each quark
propagator, in particular if $\Gamma$ couples different Dirac components. This
typically results in a relatively noisy signal\,\cite{Bulava_a2a}. By
contrast, for correlators which involve only diagonal combinations of spinor
components, it is possible to compute the two-point function stochastically
using only a single random source, which is distributed over all colour and
Dirac components within one timeslice. The relative numerical effort compared
to using a point source is thereby reduced by a factor~12 per hit: Setting
$\Nr=12$ results in the same number of inversions that must be performed,
while the statistical error can be expected to be reduced. This is the
so-called ``one-end trick'' \cite{Foster,McNeile}.

The situation in the case of correlation functions, in which different spinor
components are coupled, is less favourable but can be dealt with via the
generalized one-end trick, sometimes also referred to as the ``linked source
method'' \cite{ETM}. It amounts to choosing a spin-diagonal random source
vector, which has support only for a particular spin component~$\tau$ and a
single timeslice~$y_0$ (e.g.  $y_0=0$), viz.
\begin{equation}
  \big(\eta^{(r)}\big)^{b}_{\sigma}(y)=
  \big(\xi^{(r)}\big)^{b}(\vec{y})\,\delta_{0y_{0}}\,
  \delta_{\sigma\tau},
\end{equation}
where the components of the stochastic vector $\xi^{(r)}$ are drawn from a
distribution $\mathcal{D}$ and satisfy
\begin{equation}
   \left\langle \xi^{a}(\vec{x}) (\xi^{\dag})^{b}(\vec{y})
   \right\rangle_{\rm{src}} = \delta_{\vec{x}\vec{y}}\,\delta^{ab}.
\end{equation}
Solving the linear system of eq.~(\ref{eq:linsyst}) for spin
component $\tau$ yields the solution vector $\Phi(x)$, i.e.
\begin{equation}\label{solution}
   \big(\Phi^{(r)} \big)^{a}_{\alpha;\tau}(x) =
   \sum_{\vec{y}}\sum_{b}S_{\alpha\tau}^{ab}(x,y)
   \big|_{y_{0}=0}\xi^{b}(\vec{y}).
\end{equation}
The correlation function at zero momentum is then obtained as
\begin{equation}
 \widetilde{C}_{2}(t) =-\left\langle
 \sum_{\vec{x}}\sum_{a,\alpha,\tau}\left\langle
 \left[(\Gamma\gamma_{5}) \Phi(x)^{\dag} \right]^{a}_{\tau;\alpha}
 \left[ (\gamma_{5}\tilde{\Gamma}) \Phi(x) \right]^{a}_{\tau;\alpha}
 \right\rangle_{\rm{src}} \right\rangle\, ,
\end{equation}
which is the stochastic estimator of the two-point function at zero
momentum of eq.~(\ref{final_2pt}) after applying
eq.\,(\ref{orthonormality_condition}). The generalized one-end trick
can be applied whenever the stochastic source vector commutes with the
given choice of $\Gamma$-matrices. The spin-diagonal source vector
$\xi^{(r)}$ satisfies this requirement by construction. Compared with
the point source, the numerical effort is reduced by a factor three
per hit.

The use of linked sources is not mandatory for pseudoscalar mesons,
since the diagonal $\Gamma$-structure of the associated correlators
($\Gamma\gamma_{5}=1$) implies that stochastic noise vectors commute
without any modification. However, during the course of a simulation,
many different hadronic channels, involving arbitrary Dirac structures
at both the source and sink, are considered. In order to facilitate
the cost comparison for different correlator channels, we have
implemented linked sources by default, and below we proceed to compare
their effectiveness in simulations covering a range of light quark
masses for both the pseudoscalar and vector channel. According to
ref.\,\cite{andreas_article}, the use of linked sources for
pseudoscalar mesons was not found to be inferior compared with
non spin-diagonal sources, at fixed computational cost.


One potential drawback of the one-end trick is that two-point
functions cannot be computed for arbitrary momenta using a given set
of noise vectors. Owing to the automatic summation over the spatial
source and sink coordinates, a specific momentum is selected. To
utilize the one-end trick for computing the two-point correlator at a
given non-zero momentum, the set of noise vectors $\left\lbrace
\eta^{(r)}(y_0,\vec{y})|r=1,\dots,N_{r}\right\rbrace$ must be transformed separately for
each selected momentum with an appropriate phase $e^{i\vec{p}\vec{y}}$
prior to performing the inversion. 
%
It is worth noting that (partially) twisted boundary conditions 
\cite{partially_twisted}
can be
successfully combined with random
sources\,\cite{ETM,RBCpion}.
Again, for each momentum channel, $\Nr$ extra inversions are then
required, and thus the numerical cost increases in relation to the
point source, where the two-point function can be projected at least
on the Fourier modes at negligible additional cost. 

\section{Simulation setup}\label{Simulation_Setup} 

This work is based on gauge configurations with $\Nf=2$ flavours of
non-perturbatively $\rmO(a)$-improved Wilson fermions which have been
generated as part of the CLS effort\,\cite{CLS}, using the
deflation-accelerated DD-HMC algorithm \cite{DDHMC}. This algorithm
combines domain-decomposition (DD) methods \cite{DD} with the Hybrid
Monte Carlo (HMC) algorithm \cite{HMC} and the Sexton-Weingarten
multiple-time integration scheme \cite{Sexton}. All ensembles in this
work were generated for $\beta=5.3$, a choice for which the
coefficient of the Sheikholeslami-Wohlert term was determined as
$c_{\rm{sw}}=1.90952$\,\cite{Jansen}.


\begin{table}[tbh]
\begin{center}
\begin{tabular}{c c c c c c}
\noalign{\vskip0.3ex} \hline\hline\noalign{\vskip0.3ex} Run & Lattice
& Number cfgs. & $\kappa_{\rm{sea}}$ & $am_\pi$ & $m_\pi\,[\hbox{MeV}]$ \\
\noalign{\vskip0.3ex} \hline\noalign{\vskip0.3ex}
 $E_2$ & $64\times32^3$ & 158 & 0.13590 & 0.24292(29) & 760 \\
 $E_3$ & $64\times32^3$ & 156 & 0.13605 & 0.20645(37) & 645 \\
 $E_4$ & $64\times32^3$ & 162 & 0.13610 & 0.19305(41) & 605 \\
 $E_5$ & $64\times32^3$ & 159 & 0.13625 & 0.14345(55) & 450 \\
\hline\hline
\end{tabular}
\end{center}
\caption{\small Simulation parameters and pion masses. The latter are
  determined using the one-end-trick with $\Nr=6$ hits each on three different
  timeslices for the source vector.}
\label{tab:parameters}
\end{table}

Our simulation parameters are listed in Table\,\ref{tab:parameters}.
The ensembles contain configurations which are sufficiently separated
in Monte Carlo time such that autocorrelations can be safely ignored.
In this work we always set the valence quark mass equal to that of the
sea quarks. The inversions of the Wilson-Dirac operator of
eq.\,(\ref{eq:linsyst}) were performed using a Schwarz-preconditioned
generalized conjugate residual (SAP+GCR) algorithm \cite{GCR_LU}.
Further simulations details are described in
\cite{Mainz_lat09and10}. The conversion of pion masses into physical
units was performed using the preliminary scale determination via the
mass of the Omega baryon\,\cite{Hippel_lat11}, which yields
$a=0.063\,\fm$ at $\beta=5.3$.

Mesonic two-point correlators were computed in the pion
($\Gamma=\gamma_{5}$) and rho ($\Gamma=\gamma_{i}$) channels. For the
latter we averaged the contributions from all three spatial
$\gamma$-matrices. Only flavour non-singlet correlators were
considered. The quark propagators entering the correlation functions
were computed either using a point source or by applying a random
source in the manner of the generalized one-end trick described above.
For the latter linked sources with $\Nr=1,\,3$ and 6~hits were used. While
for $\Nr=3$ hits the numerical effort expressed in terms of the number
of inversion remains the same as for the point source, it reduces by a
factor of three for $\Nr=1$. Six hits represent twice as many
inversions compared to the point source. In order to study the scaling
of the variance with $\Nr$ in a more detailed manner, we used as many
as 25 hits for ensemble~$E_4$.
The accuracy of our calculations is enhanced by averaging the results obtained for three different
locations for each type of source, corresponding to source positions
$x_{0}/a=0,21$ and $42$.

In our analysis the forward-backward symmetry of the correlators was
exploited to average the data about the central timeslice $T/2$. In
order to extract the masses of the ground state we performed
correlated $\chi^2$-minimizing fits to the folded correlation
functions. The next-to-lowest state was explicitly taken into account,
by substituting its energy by $3m_{\pi}$ in the pseudoscalar channel
and $2\sqrt{m_{\pi}^2 + (2\pi/L)^2}$ in the vector channel,
respectively\,\cite{fit_mass}.
Statistical errors were determined via the single-elimination jackknife
method.  As will be explained below, the achieved statistical precision on
spectral quantities enters our definition of a measure for the efficiency of
stochastic sources. In order to quantify the significance of this measure we
have used Berg's proposal\,\cite{Berg92} of a nested jackknife procedure (also
referred to as second-level jackknife) to estimate the fluctuations of the
statistical error (i.e the ``error of the error'').

\section{Results: Pseudoscalar and vector meson two-point
  correlators}\label{results} 

Our objective is to compare the performance of point sources and the
generalized one-end trick when computing light meson masses. Therefore, for a
variety of different quark masses the resulting variances are monitored as a
function of the computational effort, which is expressed in terms of the
number of inversions, $N_{\rm{inv}}$, which are required to solve the linear
system of \eq{eq:linsyst} for one particular source position. This is
motivated by the observation that the number of iterations of the deflated
SAP+GCR solver does not depend on the source type but only on the simulated
quark mass. Note that $N_{\rm{inv}}$ represents the number of inversions per
source position.

\begin{figure}
\begin{center}
\includegraphics[width=14.cm]{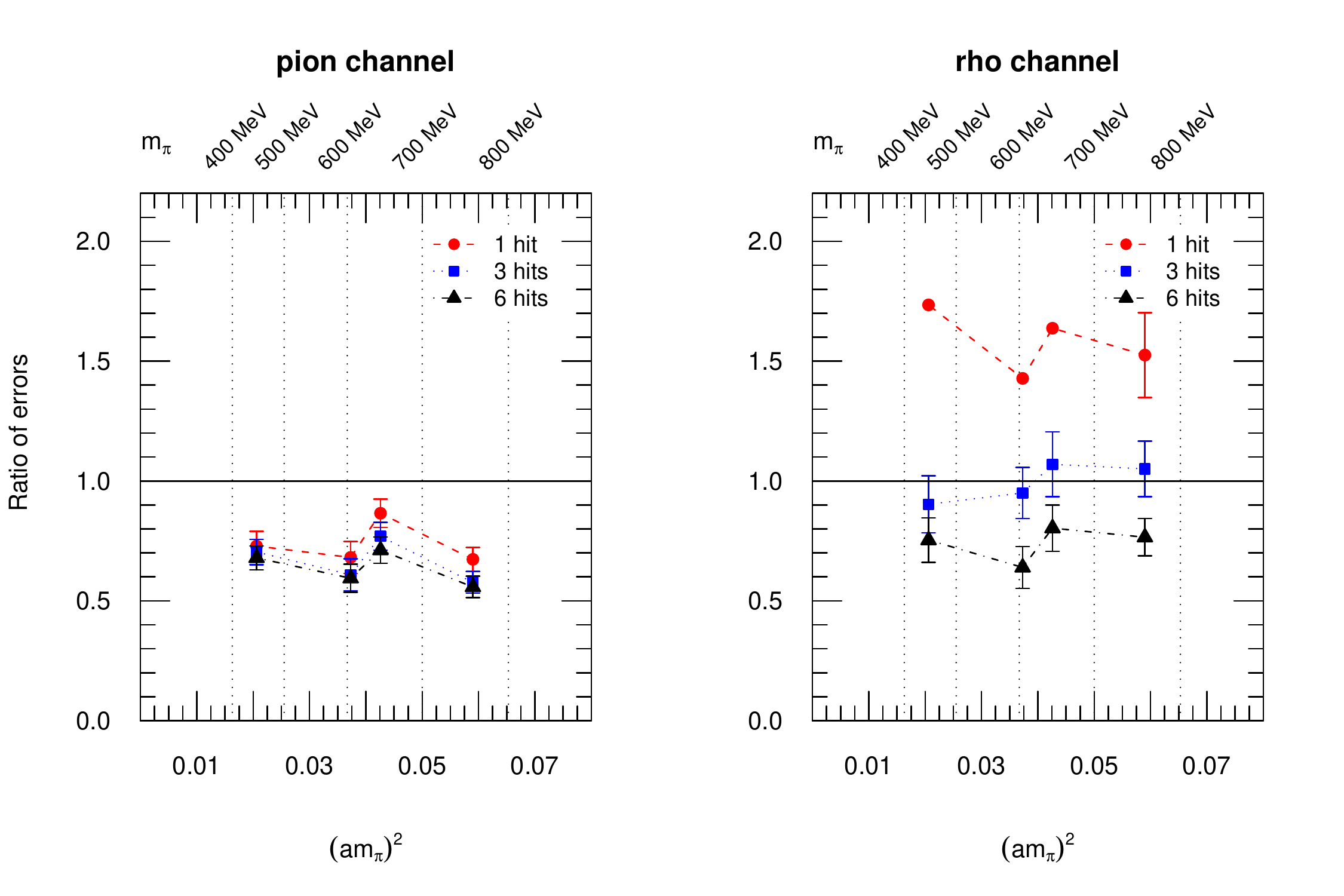}
\caption{\small The statistical error of meson masses
  computed using volume sources normalized by the error obtained with
  the point source, plotted against the squared pion mass for $\Nr=1,\,3$
  and 6 hits. Error bars (where shown) were obtained using a nested jackknife
  procedure.}
\label{fig:rel_error} 
\end{center}
\end{figure}

Table\,\ref{tab:masses} contains the results of the pion and rho meson
mass fits using both source types for the various gauge field
ensembles whose simulation parameters are listed in
Table\,\ref{tab:parameters}. By default, results from the three source
positions are averaged over. The results are illustrated in
Fig.\,\ref{fig:rel_error}, which shows the ratio of errors obtained
using stochastic volume sources and point sources, respectively.
\begin{table}[tbh]
\begin{center}
\begin{tabular}{lccccc}
\hline\hline
\multicolumn{1}{c}{Source type}& $N_{\rm{inv}}$ &$E_2$ & $E_3$ & $E_4$ & $E_5$\\\hline
                &    & \multicolumn{4}{c}{$am_{\pi}$}\\\hline
Point           & 12 &$0.24216(52)$ & $0.20647(52)$ & $0.19240(69)$ & $0.14264(81)$ \\
Volume: 1 hit   &  4 &$0.24337(35)$ & $0.20684(45)$ & $0.19335(47)$ & $0.14402(59)$\\
Volume: 3 hits  & 12 &$0.24309(30)$ & $0.20650(40)$ & $0.19320(42)$ & $0.14346(57)$\\
Volume: 6 hits  & 24 &$0.24292(29)$ & $0.20645(37)$ & $0.19305(41)$ & $0.14345(55)$\\\hline
                &    & \multicolumn{4}{c}{$am_{\rho}$}\\\hline
Point           & 12 &$0.3782(38)$ & $0.3501(49)$ & $0.3353(56)$ & $0.2842(80)$ \\
Volume: 1 hit   &  4 &$0.3819(57)$ & $0.3523(80)$ & $0.3166(80)$ & $0.2909(139)$\\
Volume: 3 hits  & 12 &$0.3843(39)$ & $0.3518(52)$ & $0.3367(54)$ & $0.2820(73)$ \\
Volume: 6 hits  & 24 &$0.3858(29)$ & $0.3559(39)$ & $0.3333(36)$ & $0.2842(61)$ \\
\hline\hline
\end{tabular}
\end{center}
\caption{\small Pion (upper half) and rho meson (lower half) masses computed using point sources
  and the generalized one-end trick with $N_{r}=1,3$ and $6$ hits for
  the ensembles $E_2-E_5$. $N_{\rm{inv}}$ denotes the number of
  inversions per source position. Fit ranges were chosen as
  $10\,{\leq}\,x_{0}/a\,{\leq}\,30$ and
  $11\,{\leq}\,x_{0}/a\,{\leq}\,30$ in the pion and rho meson
  channels, respectively. In case of the lightest rho meson the fit
  range was reduced to $11\,{\leq}\,x_{0}/a\,{\leq}\,27$, due to
  strong fluctuations around the central timeslice.}\label{tab:masses}
\end{table}

Our investigation of the uncertainty in the statistical error estimate itself
via a nested jackknife procedure revealed that the fluctuations in the error
estimate are quite small, amounting on average to about $(4-7)$\% in the pion
channel and $(7-10)$\% in the rho meson channel, respectively.

Comparing the conventional point source to the generalized one-end trick in
the pion channel we observe that random noise sources lead to a considerable
improvement of the statistical signal. Already a single random noise vector, corresponding to one third of the relative computational effort leads to a
significantly reduced variance. The variance saturates very quickly when
increasing the number of hits. This implies that the gauge noise dominates
over the stochastic noise of random volume sources.  Furthermore, a
considerable improvement of at least 25 percent at equal cost is observed for
all simulated quark masses. Our confidence in the results is supported by the
quality of plateaus of effective masses. The plots for the pion are shown
in Fig.\,\ref{fig:pion_plateau} of Appendix\,\ref{pion plateau} and illustrate that
outliers of the plateau are suppressed due to the volume averaging effect of
our stochastic sources.

\begin{figure}[pb]
\begin{center}
\includegraphics[width=14cm]{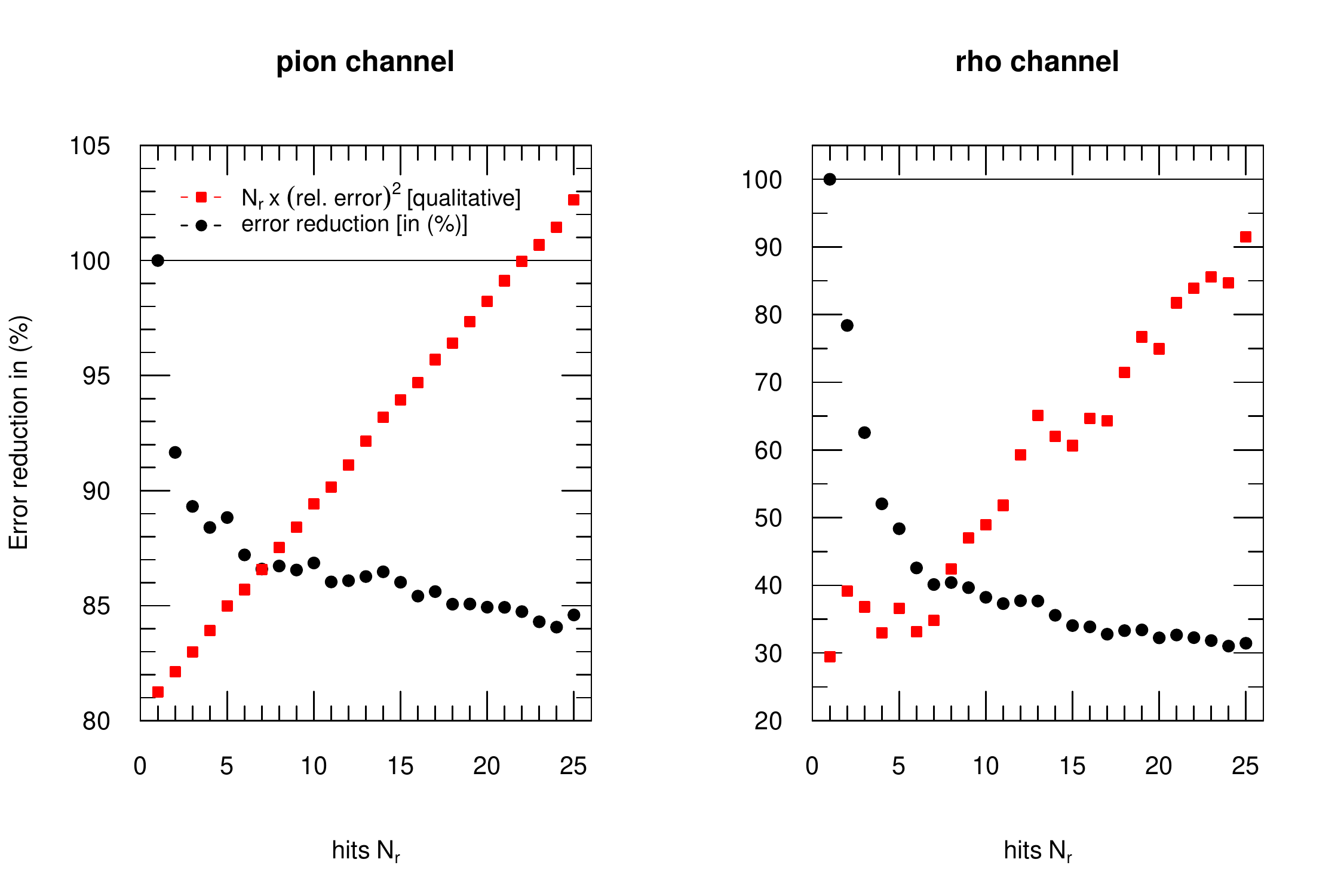}
\caption{\small Effectiveness of stochastic sources as a function of the
  number of hits, $\Nr$, for ensemble $E_4$ in the pseudoscalar (left panel) and
  vector (right panel) channels. Solid black points represent the statistical
  error relative to the single-hit volume source. Red squares denote the
  variance scaled by the number of hits.}
\label{fig:saturation} 
\end{center}
\end{figure}

In the vector channel the generalized one-end trick is less effective. A
single hit does not suffice to reach the precision of the point source method
and no reliable estimates for the errors of the error were
obtained. Therefore, in the right panel of Fig.\,\ref{fig:rel_error} the
errorbars of the results of a single hit are suppressed for small quark
masses. However, at equal computational cost the variances are comparable and,
in particular, for the two lightest quark masses the volume source slightly
gains over the conventional point source. Contrary to the pion channel, it is seen that the stochastic noise is not immediately saturated and increasing the number of hits reduces the variance significantly.\footnote{The effective mass plots of the rho meson are shown in Fig. \,\ref{fig:rho_plateau} of Appendix \,\ref{rho plateau}.} In order to study the scaling of the variance in more detail we
have computed the pseudoscalar and vector correlators on the $E_4$ ensemble
for up to 25~hits. The results are shown in Table\,\ref{tab:sat} and
Fig.\,\ref{fig:saturation}.

\begin{table}[pt]
\begin{center}
\begin{tabular}{lccc|cccc}
\hline\hline
\multicolumn{1}{c}{hits}& $N_{\rm{inv}}$
&\multicolumn{2}{c}{$E_4$}&\multicolumn{1}{|c}{hits} & $N_{\rm{inv}}$
&\multicolumn{2}{c}{$E_4$}\\\hline 
   &     &$am_{\pi}$ & $am_{\rho}$     &    &     &$am_{\pi}$ & $am_{\rho}$\\\hline
pt & 12  &$0.19240(69)$ & $0.3353(56)$    & 8  & 32  &$0.19302(41)$ & $0.3370(35)$ \\
1     &  4  &$0.19335(47)$ & $0.3166(80)$ & 9  & 36  &$0.19304(41)$ & $0.3361(34)$\\
2     &  8  &$0.19326(44)$ & $0.3352(67)$ & 10 & 40  &$0.19299(41)$ & $0.3363(33)$\\
3     & 12  &$0.19320(42)$ & $0.3367(54)$ & 13 & 52  &$0.19291(41)$ & $0.3377(32)$\\
4     & 16  &$0.19300(42)$ & $0.3288(43)$ & 16 & 64  &$0.19286(40)$ & $0.3399(29)$\\
5     & 20  &$0.19302(42)$ & $0.3328(41)$ & 19 & 76  &$0.19289(40)$ & $0.3408(29)$\\
6     & 24  &$0.19305(41)$ & $0.3333(36)$ & 22 & 88  &$0.19291(40)$ & $0.3417(28)$\\
7     & 28  &$0.19301(41)$ & $0.3346(34)$ & 25 & 100 &$0.19293(40)$ & $0.3430(27)$\\
\hline\hline
\end{tabular}
\end{center}
\caption{\small Variance reduction of the generalized one-end trick as
  a function of the number of hits. Pion and rho meson masses for
  $N_{r}=1,\dots,25$ hits computed on the ensemble $E_4$. The values
  of the point source (labeled by 'pt') are taken from
  Table\,\ref{tab:masses}.}\label{tab:sat}
\end{table}

The results compiled in the table demonstrate that the variance in the pion
channel is saturated already after performing three hits, indicating that this
correlator is dominated by the gauge noise. Increasing $\Nr$ from~3 to~25
results in a small additional reduction of the error in the pion mass of only
about 5\%.

By contrast, the stochastic noise dominates in the rho meson
channel. Increasing $\Nr$ from~1 to~7 produces a reduction of the statistical
error by~60\%, and a further 10\% can be gained if $\Nr$ is as large
as~25. Assuming that the error scales like $\mathcal{O}(1/\sqrt{N_{r}})$, one
expects that the squared error times the number of inversions $\Nr$ is
constant. As the right panel in Fig.\,\ref{fig:saturation} shows quite
clearly, this is indeed true in the vector channel for $\Nr\lesssim7$. When
larger values of $\Nr$ are considered, the rate of error reduction slows down
relative to the extra number of inversions, and thus the procedure becomes
increasingly ineffective in terms of computational overhead.

In order to formulate a more quantitative criterion for the
performance of point and stochastic sources, we define the relative
efficiency, $\epsilon_{ij}$, of two procedures $i$ and $j$, as
\begin{equation}
\epsilon_{ij} :=
 \frac{[(\hbox{variance}) \times (\hbox{number of inversions})]_{i}}
      {[(\hbox{variance}) \times (\hbox{number of inversions})]_{j}}
 -1.
\label{eq:rel_cost}
\end{equation}
Thus, the ratio of the squared error is scaled with the ratio of the
computational cost. In Table\,\ref{tab:releff} we list the values for
$\epsilon_{ij}$ in the vector channel for ensemble~$E_4$. The information in
this matrix-like table must be interpreted in the following way: a negative
entry~$\epsilon_{ij}$ in row~$i$ and column~$j$ means that procedure~$i$ is
more efficient than~$j$ by $(100\times|\epsilon_{ij}|)\,\%$. From the numbers in
Table\,\ref{tab:releff} one concludes that, in the vector channel, a volume
source with $\Nr=3$ is only slightly more efficient than the point source,
even though the number of inversions performed in both cases is the
same. However, if one demands that the statistical error be at least as small
as for the point source, a larger number of hits is more favourable: According
to Table\,\ref{tab:sat}, $\Nr=6$ seems to be the optimal choice, since the
extra numerical effort is more than compensated by the resulting reduction in
the variance.

\begin{table}
\begin{center}
\begin{tabular}{c| c c c c c}
\hline\hline
           & point & $\Nr=3$ & $\Nr=6$ & $\Nr=9$ & $\Nr=16$ \\
\hline
point    & $0$     & $0.08$   & $0.21$   & $-0.10$  & $-0.30$   \\
$\Nr=3$  & $-0.07$ & $0$      & $0.13$   & $-0.16$  & $-0.35$   \\
$\Nr=6$  & $-0.17$ & $-0.11$  & $0$      & $-0.25$  & $-0.42$   \\
$\Nr=9$  & $ 0.11$ & $0.19$   & $0.34$   & $0$      & $-0.23$   \\
$\Nr=16$ & $ 0.43$ & $0.54$   & $0.73$   & $0.29$   & $0$  \\
\hline\hline
\end{tabular}
\caption{The relative efficiency $\epsilon_{ij}$ in the vector channel
  for ensemble~$E_4$.} 
\label{tab:releff}
\end{center}
\end{table}

\begin{figure}
\begin{center}
\includegraphics[width=14cm]{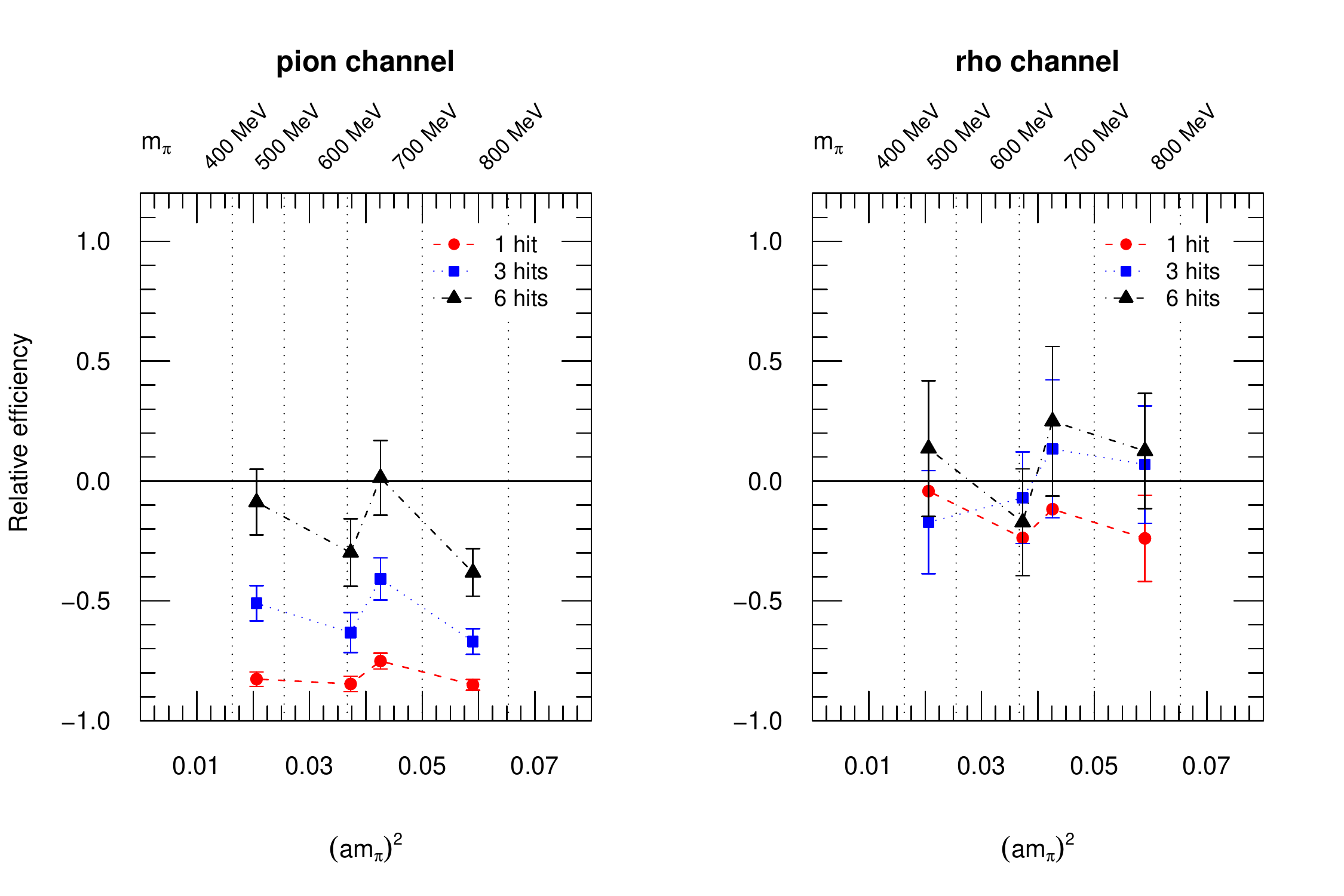}
\caption{\small Relative efficiency as defined in eq.\,(\ref{eq:rel_cost})
  with respect to the point source as a function of the squared pion mass for
  the pion (left panel) and rho meson (right panel) using $\Nr=1,3$ and 6
  hits.} \label{fig:rel_cost}
\end{center}
\end{figure}

In Fig.\,\ref{fig:rel_cost} the relative efficiency defined in
eq.\,(\ref{eq:rel_cost}) with respect to the point source is plotted as a
function of the squared pion mass. For a single hit the volume source is about
a factor of five more efficient than the point source in the pion channel
whereas at identical computational cost the volume source still outperforms
the point source by a factor of two. As pointed out before, this loss of
efficiency is due to the fact that in the pion channel the variance is
completely dominated by the gauge noise. The gain in the vector channel is not
so obvious. A stochastic volume source with $\Nr=6$ hits appears to be a good
compromise between numerical effort and statistical accuracy across the entire
mass range under study, despite the large uncertainties in the relative
efficiency $\epsilon_{ij}$ in the vector channel (see
Figs.\,\ref{fig:rel_error} and\,\ref{fig:rel_cost}).

\begin{figure}
\begin{center}
\includegraphics[width=10cm]{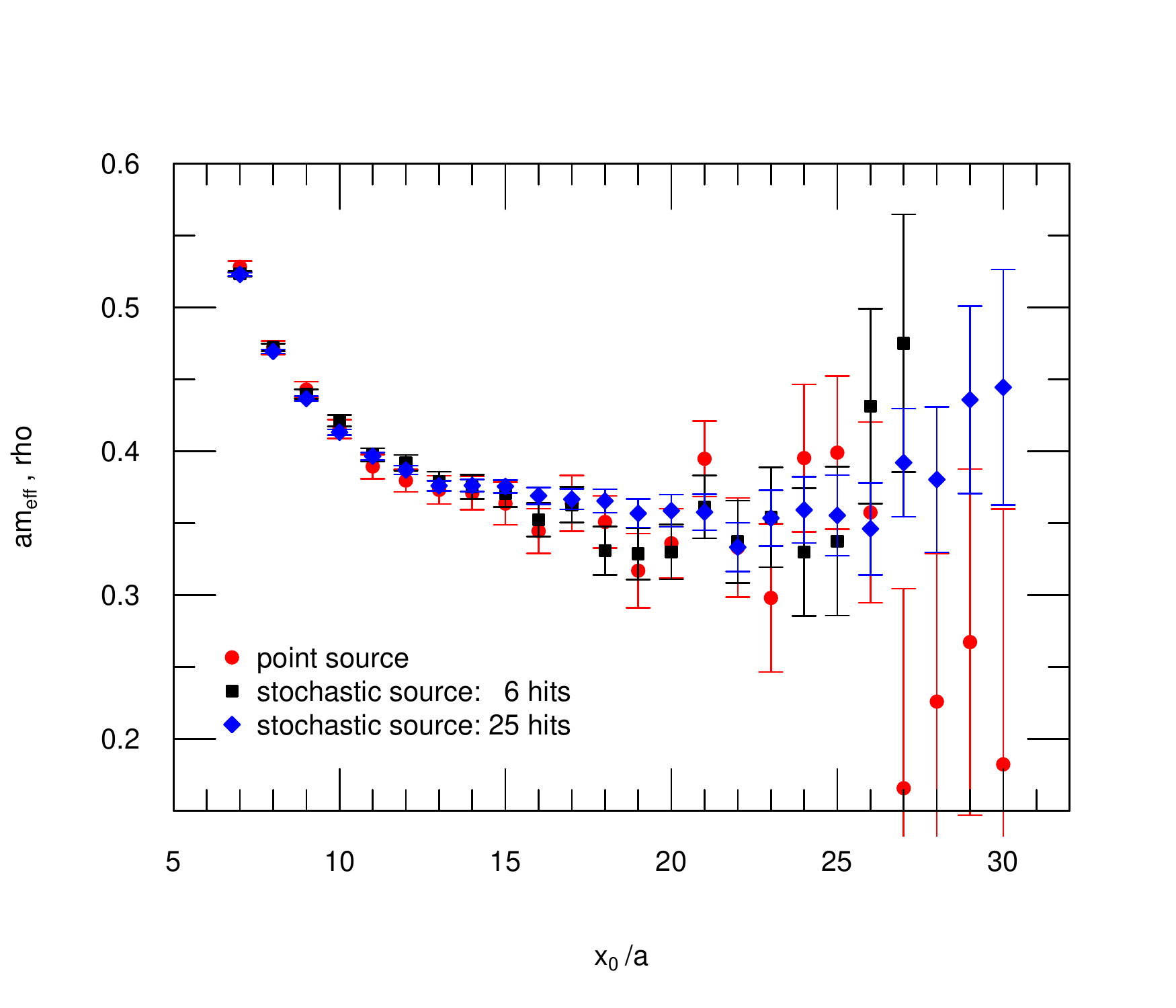}
\caption{\small Effective masses in the vector channel obtained by means of the point
  source and random sources with~6 and~25 hits.} \label{fig:effmrho}
\end{center}
\end{figure}

Another beneficial effect of using random sources in the vector channel can be
seen from the effective mass plot shown in Fig.\,\ref{fig:effmrho}. By
comparing the data obtained using $\Nr=6$ and
$\Nr=25$ hits with those of the point source, one clearly sees that not only the statistical error decreases for a
large hit number but that the overall quality of the plateau is much improved
as well. This should make for much more reliable estimates of the mass in the
vector channel.

Finally we investigate the effects on the variance of distributing random noise sources on several timeslices instead of placing them on a single one.
 Let $N_{\rm{tot}}$ denote any
number of inversions of the Dirac operator. If the random noise source is
placed on $N_{\rm{ts}}$ different timeslices, and if $\Nr$ hits are performed
for each of these source positions, then $N_{\rm{tot}}=N_{\rm{ts}}{\cdot}\Nr$. Can the
variance be reduced by choosing $N_{\rm{ts}}$ and $\Nr$ such that
$N_{\rm{tot}}$ stays fixed?

\begin{table}
\begin{center}
\begin{tabular}{c c c  | c c  | c c  }
\hline\hline
&  $\Nr$  &  $N_{\rm{ts}}=1$ [0]     & $\Nr$ &
  $N_{\rm{ts}}=2$ [0,21]  & $\Nr$ &$N_{\rm{ts}}=3$
  [0,21,42]  \\  
\hline
&3 & 0.19327(58)   &   &               & 1 & 0.19335(47)\\
&6 & 0.19315(55)   & 3 &  0.19330(49)  & 2 & 0.19326(44)\\
&9 & 0.19303(54)   &   &               & 3 & 0.19320(42)\\
$am_{\pi}$&12& 0.19296(53)   & 6 &  0.19314(47)  & 4 & 0.19300(42)\\
&15& 0.19297(54)   &   &               & 5 & 0.19302(42)\\
&18& 0.19304(54)   & 9 &  0.19306(46)  & 6 & 0.19305(41)\\
&21& 0.19307(54)   &   &               & 7 & 0.19301(41)\\
&24& 0.19308(54)   & 12 & 0.19293(46)  & 8 & 0.19302(41)\\
\hline 
&3  & 0.3290(89)  &     &            & 1  & 0.3166(80)\\
&6  & 0.3268(59)  & 3   & 0.3303(59) & 2  & 0.3352(67)\\
&9  & 0.3259(50)  &     &            & 3  & 0.3367(54)\\
$am_{\rho}$&12 & 0.3285(48)  & 6   & 0.3265(42) & 4  & 0.3288(43)\\
&15 & 0.3329(45)  &     &            & 5  & 0.3328(41)\\
&18 & 0.3366(44)  & 9   & 0.3308(40) & 6  & 0.3333(36)\\
&21 & 0.3387(45)  &     &            & 7  & 0.3346(34)\\
&24 & 0.3429(44)  & 12  & 0.3349(36) & 8  & 0.3370(35)\\
\hline\hline
\end{tabular}
\caption{\small Comparison of effective masses in the pion (upper half) and rho (lower half) channels,
  averaged over $N_{\rm{ts}}$ different source positions which are indicated in the square brackets.
  $\Nr$ denotes the
  performed number of hits per source position. The total number of
  inversions, $N_{\rm{tot}}=N_{\rm{ts}}{\cdot}\Nr$, is constant in each row of
  the table.}
\label{tab:ts_vs_hits}
\end{center}
\end{table}

The results of such an analysis are shown in Table~\ref{tab:ts_vs_hits}. Each
row represents a particular fixed value of $N_{\rm{tot}}$. In the pion channel
there are clear indications that the statistical error decreases when more
timeslices are used. In the vector channel this effect is less pronounced but
becomes evident when $N_{\rm{ts}}{\cdot}\Nr\;\gtaeq\;12$. Thus, a good strategy
to enhance that statistical accuracy of mesonic two-point correlation
functions is to use as many timeslices as one can afford for a fixed total
number of inversions.

\newpage

\section{Summary and conclusions}\label{summary}

In this paper we have presented a systematic study of random noise sources for
the calculation of mesonic two-point correlation functions, using $O(a)$
improved Wilson fermions as our discretization. Specifically, we have
investigated the effectiveness of the generalized one-end trick in the
pseudoscalar and vector channels, for pion masses ranging from~450 to
760\,\MeV. Our spatial volumes correspond to a box length of $L=2\,\fm$. The
total number of inversions of the lattice Dirac operator serves as our measure
for the computational cost.

Our findings are best summarized by a list of empirical observations:
\begin{itemize}
\item At equal computational cost, stochastic volume sources significantly
  enhance the statistical accuracy of correlation functions in the
  pseudoscalar channel. Here the signal is dominated by fluctuations in the
  gauge configurations, and the extra noise introduced by the stochastic
  procedure is rapidly suppressed.
\item Stochastic volume sources help to stabilize mass estimates in the vector
  channel. In addition to reducing the variance, stochastic volume sources
  also produce a more pronounced plateau in the effective mass plot. Compared
  to the pion channel, however, a larger number of hits must be performed,
  before the improvement is clearly visible. In order to observe a clear
  advantage over ordinary point sources in our studied mass range, the
  numerical effort must at least be doubled for the generalized one-end trick.
\item For fixed numerical cost, the quality of mesonic two-point correlation
  functions computed using random volume sources can be improved by averaging
  over several source positions, $N_{\rm{ts}}>1$, and adjusting the number of hits, $\Nr$, such as to keep
  the total number of inversions fixed. In other words,
  averaging over more source positions is more effective in reducing the variance than increasing the
  number of hits.
\end{itemize}

The use of random volume sources is an attractive method, designed to extract
more information on hadronic properties from the full propagator matrix at
reasonable cost. It is particularly useful if the number of available gauge
configurations is relatively small.
Given the convincing performance reported here for the simple case of
mesonic two-point functions, we have employed random noise sources in our
projects to determine the electromagnetic pion form factor and the form
factors for $K_{\ell3}$ decays\,\cite{Mainz_FFs}.

\section*{Acknowledgments}
We thank our colleagues participating in the CLS project for sharing gauge
ensembles. Calculations of correlation functions were performed on the
''Hydra'' cluster at the IFT, University of Madrid and the dedicated QCD
platform ''Wilson'' at the Institute for Nuclear Physics, University of
Mainz. E.E. is supported by the Research Executive Agency (REA) of the
European Union under Grant Agreement PITN-GA- 2009-238353 (ITN STRONGnet).

\newpage
\begin{appendix}
\section{Effective mass plots: pseudoscalar channel}\label{pion plateau}
\begin{figure}[h]
\begin{center}
\includegraphics[width=14cm]{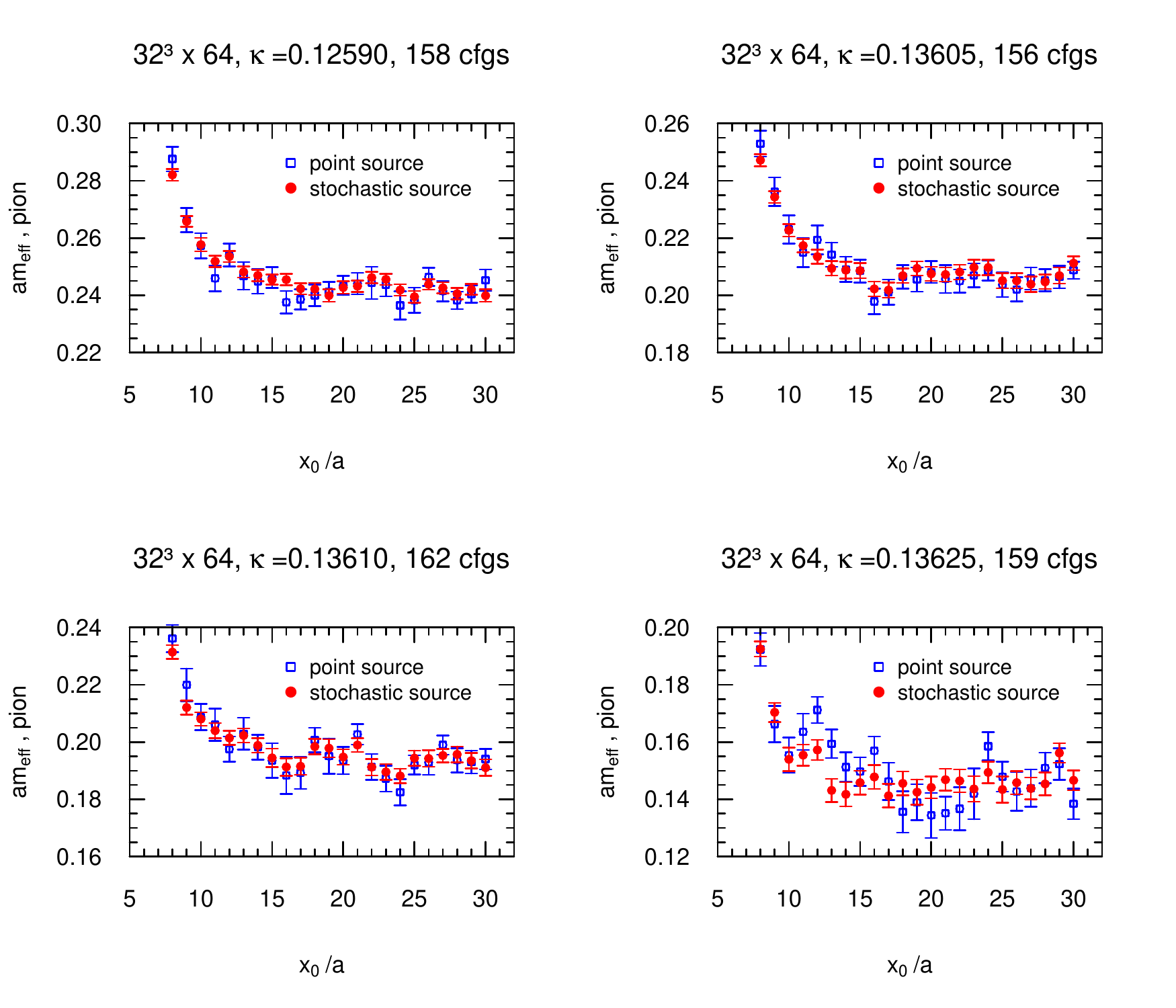}
\caption{\small Effective pion mass plots for the ensembles $E_{2}-E_{5}$
  (from top left to bottom right). Illustrated are the results for the
  point source (blue squares) and the stochastic volume source (filled
  red cirlces) at fixed cost, i.e. for $N_{r}=3$ hits.}
  \label{fig:pion_plateau}
\end{center}
\end{figure}
\newpage
\section{Effective mass plots: vector channel}\label{rho plateau}
\begin{figure}[h]
\begin{center}
\includegraphics[width=14cm]{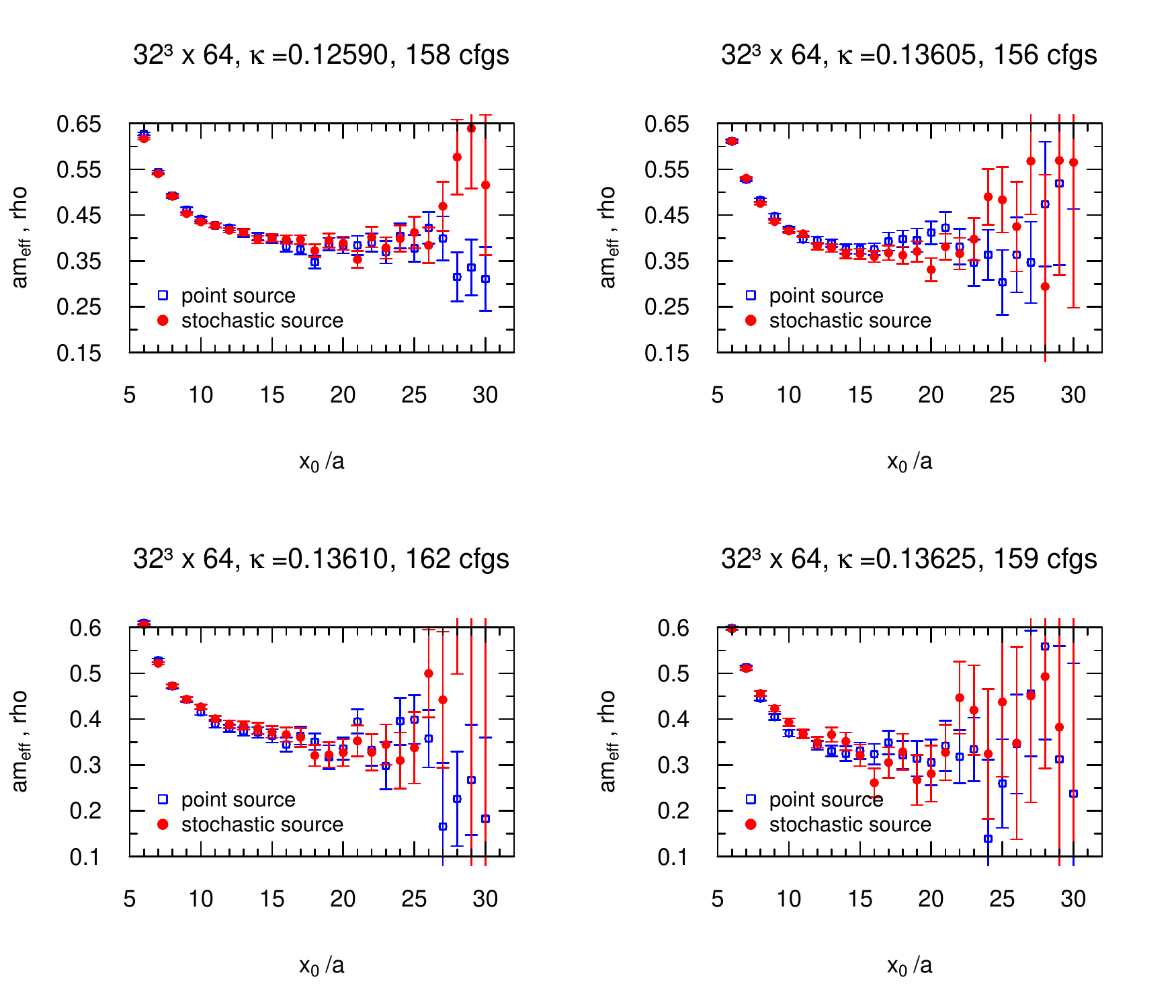}
\caption{\small Effective rho meson mass plots for the ensembles $E_{2}-E_{5}$
  (from top left to bottom right). Illustrated are the results for the
  point source (blue squares) and the stochastic volume source (filled
  red cirlces) at fixed cost, i.e. for $N_{r}=3$ hits.}
\label{fig:rho_plateau} 
\end{center}
\end{figure}  
\end{appendix}

\end{document}